\documentclass[manuscript]{emulateapj}

\newcommand{\dxdy}[2]{{\frac{\partial{#1}}{\partial{#2}}}}

\def\rhobar{{\bar\rho}}

\def\gbar{{\bar g}}

\def\nubar{{\bar\nu}}

\def\rhat{{\hat{\bf r}}}

\def\omegavec{{\bf \Omega}}
\def\grad{{\bf \nabla}}
\def\div{{\bf \nabla} \cdot}
\def\curl{{\bf \nabla} \times}
\def\vvec{{\bf v}}

\def\Bvec{{\bf B}}

\def\eten{e_{ij}}
\def\dten{\delta_{ij}}

\usepackage{graphicx}
\usepackage{amssymb}
\usepackage{bm}

\shorttitle{MHD of HD209458b}
\shortauthors{Rogers}

\begin{document}


%\title{Angular Momentum Transport by Internal Gravity Waves in Hot Stars:  Application to the Obliquities of Hot Jupiters}
\title{Magnetohydrodynamic simulations of the atmosphere of HD209458b}
\author{T.M. Rogers}
\affil{Department of Planetary Sciences, University of Arizona,
    Tucson, AZ, 85721}
\email{tami@lpl.arizona.edu}
\author{A.P. Showman}
\affil{Department of Planetary Sciences, University of Arizona,
    Tucson, AZ, 85721}
\email{showman@lpl.arizona.edu}
\begin{abstract}
We present the first three-dimensional (3D) magnetohydrodynamic (MHD) simulations of the atmosphere of HD209458b which self-consistently include reduction of winds due to the Lorentz force and ohmic heating.  We find overall wind structures similar to that seen in previous models of hot Jupiter atmospheres, with strong equatorial jets and meridional flows poleward near the dayside and equatorward near the night.  Inclusion of magnetic fields slows those winds and leads to ohmic dissipation.  We find wind slowing ranging from 10-40\% for reasonable field strengths.  We find ohmic dissipation rates $\sim 10^{17}$W at 100 Bar, orders of magnitude too small to explain the inflated radius of this planet.  Faster wind speeds, not achievable in these anelastic calculations, may be able to increase this value somewhat, but likely wont be able to close the gap necessary to explain the inflated radius.  We demonstrate that the discrepancy between the simulations presented here and previous models is due to inadequate treatment of magnetic field geometry and evolution.  Induced poloidal fields become much larger than those imposed, highlighting the need for a self-consistent MHD treatment of these hot atmospheres.  

\end{abstract}


\section{Introduction}
The field of extrasolar planet research is growing rapidly with more than 1000 planets found (as of November, 2013).  Many of these objects are large and/or close to their host star.  These large extrasolar planets have come to be known as extra-solar giant planets (EGPs) and those in close proximity to their host stars are often referred to as "hot Jupiters".  Many of these planets are observed to transit their host stars as viewed from Earth.  Combined with radial velocity data, this allows calculation of the planetary mass, radius and surface gravity.  And because they are likely synchronously rotating, planetary rotation rate can also be inferred.  

Because these close-in planets are likely tidally locked they have permanent day and night sides.  Such asymmetric heating drives strong winds in the planetary atmosphere which have been modeled by several groups \citep{coop05,iandd08,show09,rm10,lewis10,heng11}.  While these models vary in their approach and complexity, they have several common features, such as transonic wind speeds and eastward equatorial jets.  The eastward equatorial jet implies that the hottest point would be advected eastward of the substellar point, as was first predicted by \cite{show02}, and which was subsequently observed on HD 189733b by \cite{knutson07,knutson12}.  Although discrepancies exist, it appears that atmospheric circulation models with varying forms of radiation transport are doing well at describing the large scale dynamics of these planets' atmospheres.  

However, there are still several unexplained observations.  First, it was observed early on that many EGP's, particularly those that are the most strongly irradiated, have radii much larger than that expected from standard evolutionary models \citep{bod01,bod03,guillot02,ba03,laughlin05}.  The most likely solution to this observation is that there is an additional heat source in the planetary interior that slows gravitational contraction causing the radius to appear larger than expected from standard evolutionary models.  Many heating mechanisms have been proposed such as tidal heating \citep{bod01,bod03}, downward transport of mechanical energy by atmospheric circulation via waves or advection \citep{guillot02}, atmospheric turbulence \citep{youdin10} and ohmic dissipation \citep{bat10,perna10b,laughlin11}.  

Second, observations of infrared and visible light curves have been able to constrain how efficiently the planetary atmosphere transfers heat from the irradiated dayside to the cooler nightside by measuring the phase variation over the planetary orbital period.  The picture emerging is that these planets likely have a variety of atmospheric circulation patterns, with varying recirculation efficiency \citep{cowan11,pbs13}.  Observations of HD 209458b and 51 Peg \citep{cowan07}, HD189733b \citep{knutson07,knutson09a} and HD 149026b \citep{knutson09b} indicate that these planets have efficient recirculation from their day to night side and hence, show small ($<500$K) temperature variations.  However, observations of Ups And b \citep{crossfield10}, HD 179949 \citep{cowan07}, HAT-P-7b \citep{borucki09}, WASP-12b \citep{cowan12} and WASP-18b \citep{maxted13} indicate weak recirculation and hence, large day-night temperature variations.  It has been argued that a strong magnetic field could prevent efficient circulation from the day to night side by opposing such flow.  

The effects of the magnetic field are therefore, speculated to be (at least) twofold.  First a planetary magnetic field could slow day-night winds causing reduced circulation efficiencies and hence, larger day-night temperature variations \citep{perna10a,menou12,rm13}.  Second, ohmic dissipation associated with the magnetic field could lead to the heating required to explain the inflated radii of many hot Jupiters \citep{bat10,bat11,perna10b}.  These pioneering studies demonstrate the possible influence of magnetic effects on the atmospheric winds and quantify the possible behavior of the ohmic dissipation.  However, these early studies did not solve the full MHD equations.  Rather, they adopt a kinematic approach where the Lorentz force is estimated from the winds using a prescribed dipole magnetic field and assuming the induced fields generated by electrical currents can be neglected.  This approach is a natural first step, but is only valid at magnetic Reynolds numbers $\leq 1$  ($R_{m}=UL/\eta$, where $U$ is a typical speed and $L$ is a typical length scale and $\eta$ is the magnetic diffusivity).  However, the actual $R_{m}$ in these planetary atmospheres often exceed unity, particularly for the hotter planets.  This indicates the need for a fully self-consistent MHD treatment.

An exception to this prescription is the work by \cite{bat13}, which solves the MHD equations, but in the Boussinesq approximation, so they are effectively two-dimensional and they omit ohmic heating.  Here, we present fully 3D MHD simulations including the effects of compressibility and ohmic heating.   

\section{Numerical Method}
\begin{figure}
  \centering
  \includegraphics[trim=160 0 200 0,clip,width=0.95\columnwidth]{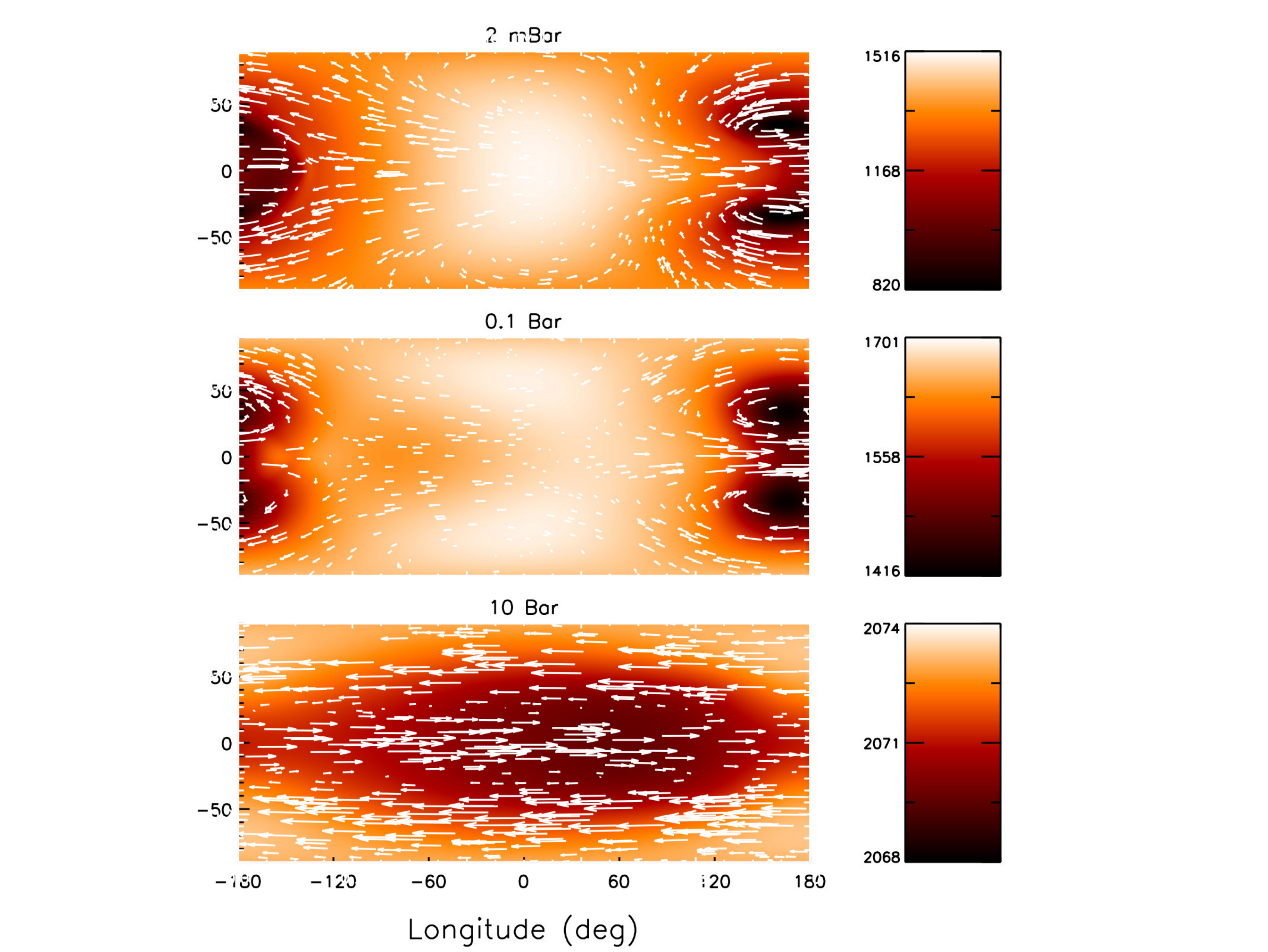}
  \caption{Winds (shown as arrows) and temperature (colors) in our
    purely hydrodynamic models at three different radii within our
    simulations after 200 rotation periods.}
\end{figure}

\begin{figure*}
  \centering
  \includegraphics[trim=0 100 0 100,clip,width=0.95\textwidth]{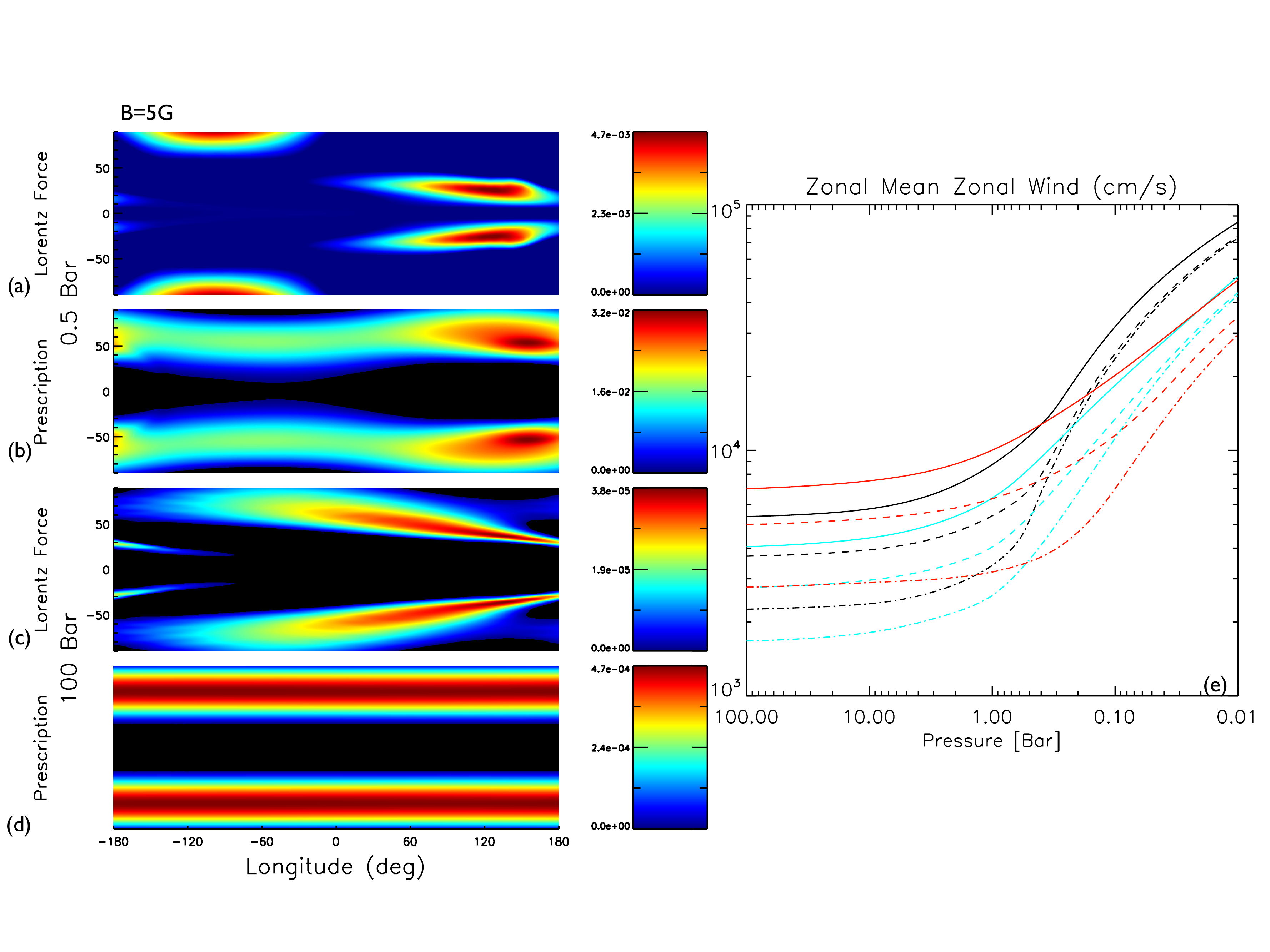}
  \caption{Effect of magnetic fields on winds.  (a) Lorentz force in
    the simulation (last term on RHS of Equation 3), (b) Magnetic drag
    calculated using Equation (6) with a constant magnetic field
    strength, 5G, at 0.5Bar and 100Bar (c,d).  The net effect of
    magnetic fields on zonal winds is shown in (e) with solid lines
    representing hydrodynamic models, dashed lines representing 5G
    fields and dot-dashed lines representing 15G fields.  Black lines
    represent equatorial flows, cyan represents mid latitude flows and
    red represents high latitude flows.}
\end{figure*}

\begin{figure*}
  \centering
  \includegraphics[trim=100 0 150 0,clip,width=0.8\textwidth]{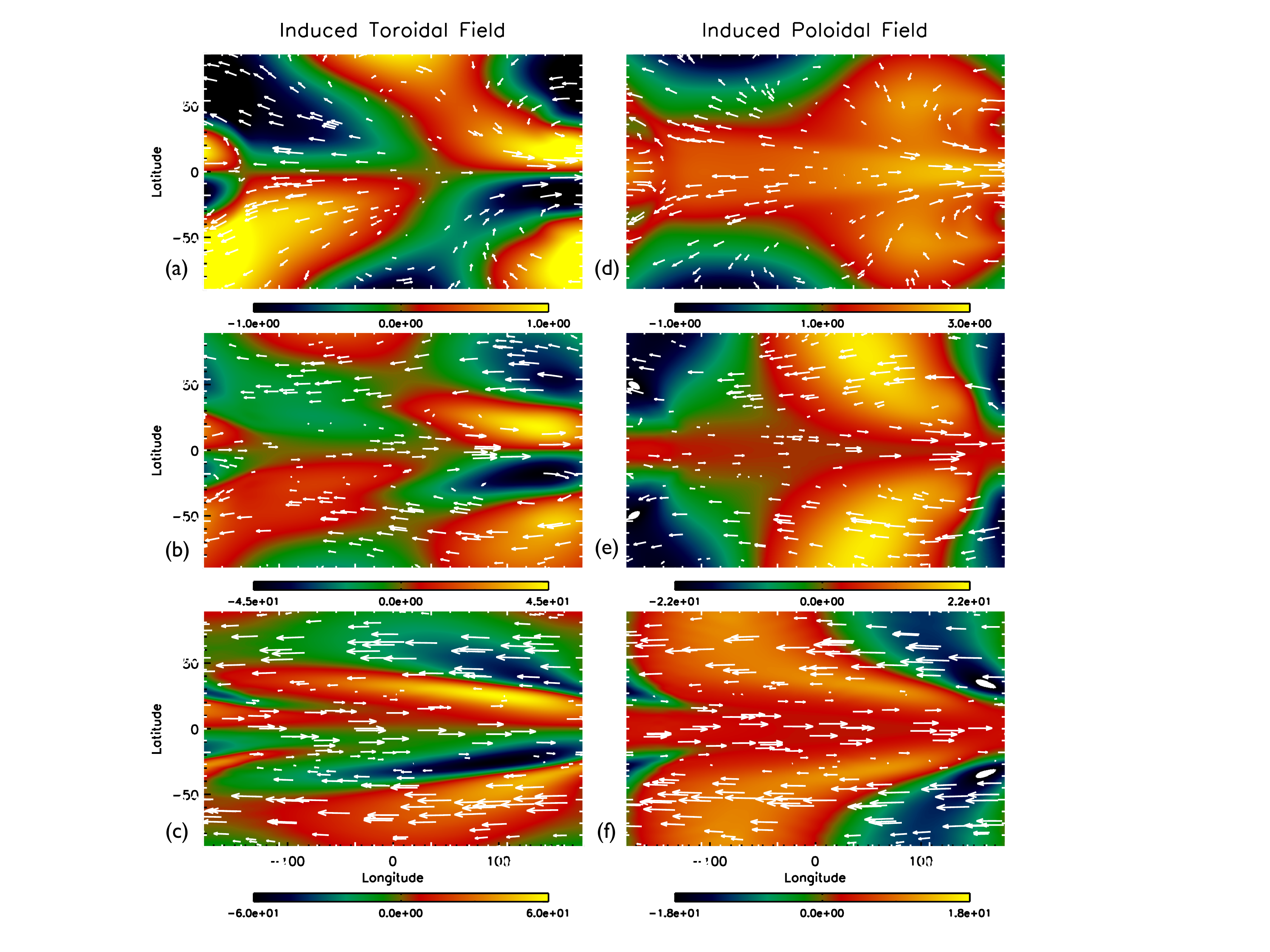}
  \caption{Azimuthal and latitudinal components of the magnetic field
    at 10 Bar (c,f), 1 Bar (b,e) and 10 mBar (a,d).  Wind speeds are
    shown as arrows.  Note amplitudes of the latitudinal field are
    $\sim$4 times the imposed value.}
\end{figure*}

\begin{figure*}
  \centering
  \includegraphics[trim=0 70 0 50,clip,width=0.95\textwidth]{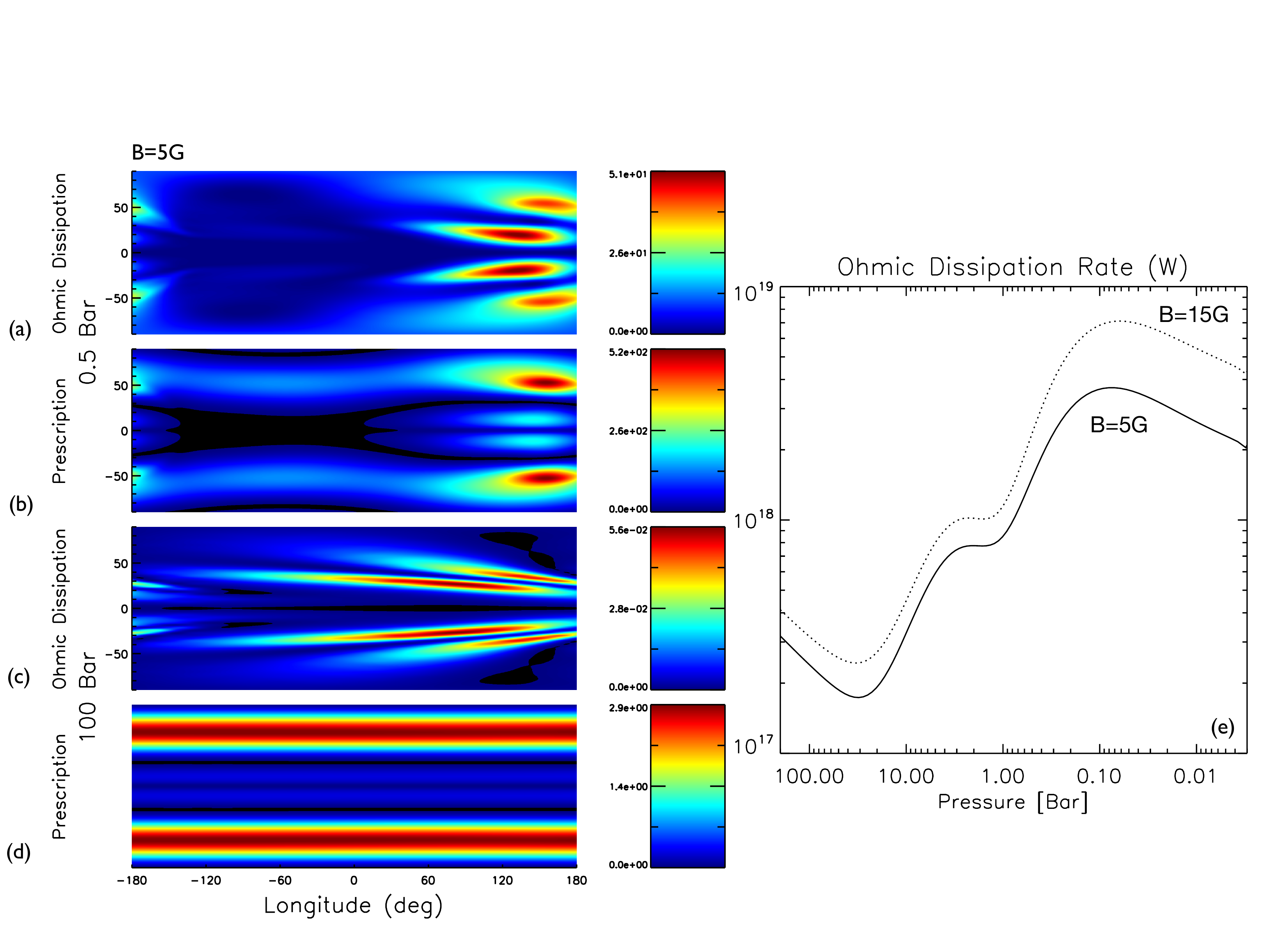}
  \caption{Ohmic dissipation within the MHD simulation, measured as
    the last term on the RHS of Equation (5) at 0.5 and 100Bar (a,c)
    and ohmic dissipation calculated using the prescription in
    Equation (6) with a constant 5G magnetic field strength (b,d).  In
    (e) we show the ohmic dissipation rate within a pressure scale
    height (in Watts), as a function of pressure.  This appears too
    weak to explain the inflated radius of this planet.}
  \label{fig:ohmd}
\end{figure*}

We use a 3D, MHD model in the anelastic approximation.  The model is based on the Glatzmaier dynamo code \citep{glatz84,glatz85}, but has some differences in the actual equations solved, the discretization and the implementation.  The model decomposes the magnetic and mass flux into a poloidal-toroidal decomposition and it is discretized using spherical harmonics in the horizontal and a fourth-order finite difference scheme in the vertical.  A description of the axisymmetric version of this code can be found in \cite{rogers11}.  The 3D version has been benchmarked with several other dynamo codes \citep{christ01,jones11}.  The model solves the following equations:   
\begin{eqnarray}
\div \rhobar \vvec &=& 0 \\
\div \Bvec &=& 0\\
\rhobar {\partial \vvec \over \partial t} +\div(\rhobar \vvec \vvec) &=& \nonumber
- \grad p - \rho \gbar \rhat + \frac{1}{\mu_0} ( \curl \Bvec ) \times \Bvec + 
\\  &&\hspace{-17mm}\mbox{} + 2 \rhobar \vvec \times \omegavec
+ \div \left[ 2 \rhobar \nubar \left(\eten - {1 \over 3} \dten \div \vvec \right)\right]\\
\frac{\partial \Bvec}{\partial t} &=&
\curl ( \vvec \times \Bvec )- \curl (\eta \curl \Bvec)
\\
\dxdy{T}{t}+(\bf{v}\cdot\bf{\nabla}){T} &=& \nonumber
-v_{r}\left[\dxdy{\overline{T}}{r}-(\gamma-1)\overline{T}h_{\rho}\right]+ \\\nonumber
&&\hspace{-25mm} \mbox{}+(\gamma-1)Th_{\rho}v_{r}
+\gamma\overline{\kappa}\left[\nabla^{2}T+(h_{\rho}+h_{\kappa})\dxdy{T}{r}\right]+ \\
&&\mbox{}+ \frac{T_{eq}-T}{\tau_{rad}}+\frac{\eta}{\mu_{o}\rho c_{p}}|\nabla \times {\bf B}|^{2}
\end{eqnarray}
Equation (1) represents the continuity equation in the anelastic approximation \citep{go69,rg05}.  This approximation allows some level of compressibility by allowing variation of the reference state density, $\rhobar$, which varies in this model by five orders of magnitude.   Equation (2) represents the conservation of magnetic flux.  Equation (3) represents conservation of momentum including Coriolis and Lorentz forces, Equation (4) represents the magnetic induction equation and Equation (5) represents the energy equation.  All variables take their usual meaning and details can be found in \cite{rogers11}, with the following few exceptions.  The magnetic diffusivity, $\eta$, is a function of temperature, as in \cite{perna10a}, but, in this first approach, we take it to be only a function of the reference state temperature (and therefore, only a function of radius).  Strong stellar irradiation is treated using a Newtonian radiative scheme as in \cite{coop05}.  This prescription is indicated by the fourth term on the right hand side (RHS) of Equation (5).  The last term of Equation (5) represents ohmic heating.   We note that these equations are non-hydrostatic and the diffusion coefficients in Equation (3) and (5) are explicit with no hyper-diffusion employed.  These values are set as low as possible while maintaining numerical stability\footnote{For reference, the Ekman number in these simulations is $1.6\times 10^{-6}$}.  We resolve the radial extent from ~$300-2\times 10^{-3}$Bar.  The resolution of the models presented is 64 x 128 x 100 ($N_{\theta}$ (latitude) x $N_{\phi}$ (longitude) x $N_{r}$, or T42 with 100 radial levels) and the code is parallelized using Message Passing Interface (MPI).  

We initially run purely hydrodynamic models for $\sim 250$ rotation periods ($P_{rot}$), then add a magnetic field.  We ran two models varying the reference state (and equilibrium) temperature:  1) the model for HD209458b from \cite{iro05} and 2) the \cite{iro05} model with 300K added at every radial level\footnote{This is clearly a crude prescription of a reference state model, however, in terms of temperature within the atmosphere, it is not significantly different than the analytic prescription by \cite{guillot10}.}.  These represent our ``cool'' and ``hot'' models, respectively (in our hot model temperatures vary from 2100K at depth to 1300K aloft) and we adopt a planetary radius, rotation rate and gravity of $9\times 10^{9}cm$, 3 days and $900 cm/s^{2}$.  We also present two different magnetic field strengths: A) A model with a 5G dipolar field at the base of the atmosphere and 3.5G at the top of the atmosphere initially and B) A higher magnetic field strength of 15G at the bottom and 10G at the top initially.  These values correspond to the radial field strength at the pole or the latitudinal field at the equator, and the radial structure is consistent with a dipolar field which falls off as $r^{-3}$.  The magnetic field strength at the base of the atmosphere is held fixed and matched to a potential field at the top of the domain.  The magnetised models are then run for $~1500 P_{rot}$ and are in a steady state.   We note at the outset, that the \cite{iro05} model is too cool to show any appreciable magnetic effects and so the results displayed in Section 3 are those for the hotter model.

\section{Results}
Figure 1 shows the temperature structure along with wind patterns within our purely hydrodynamic simulation.  We see that these models recover some of the typical features of radiatively forced atmospheric models: azimuthal flow marked by a prograde equatorial jet at most longitudes (and on zonal average), predominantly retrograde flow at higher latitudes with meridional flow towards the poles near the dayside and returning toward the equator at the nightside.  The chevron shaped patterns are typical of the Kelvin/Rossby wave forced flows described in \cite{showpolv11}.  However, our flows have lower speeds in general, peaking around a couple km/s and we note that at depth, our pattern of temperature fluctuations deviates from previous models.  This is likely due to our treatment of thermal forcing and diffusion.  We are currently looking into this issue, but note that it does not affect our results (the velocity patterns are similar to previous models and these temperature fluctuations are only a few K).  

\subsection{Slowing winds: The Lorentz Force and Magnetic Drag}
Previous work has indicated that a planetary scale magnetic field of predominantly dipolar structure could slow winds and diminish day-night circulation \citep{perna10a,menou12,rm13}.  Because previous works did not solve MHD equations this physical concept was introduced in the form of a drag term on the zonal flow equation which was proportional to the zonal flow speed divided by a magnetic timescale that represents the time required for the field to halt the zonal flow.  An order of magnitude estimate of this timescale is:
\begin{equation}
\tau_{mag}\sim\frac{\rho v_{\phi}c}{j_{\theta}\times B} \sim \frac{4\pi\eta}{B^{2}|sin\theta|}
\end{equation}
the Lorentz force is then calculated as $-v_{\phi}/\tau_{mag}$ and the ohmic dissipation rate is $v_{\phi}^{2}/\tau_{mag}$, where $v_{\phi}$ is the zonal flow speed, $j_{\theta}$ is the latitudinal current, B is the imposed dipolar field strength, $\theta$ is the latitude, and c is the speed of light.

In Figure 2 we show the azimuthal component of the Lorentz force calculated in our simulation (a,c), last term on RHS of Equation 3) and the drag term using the prescription described in Equation (6) with a constant field strength (as done by previous authors b,d) at 0.5Bar and 100Bar.  There are several important differences to note.  First, our MHD simulation predicts a Lorentz force whose peak value is an order of magnitude smaller the peak value given by the prescription encompassed in Equation (6) and second, the actual Lorentz force is more localized spatially than the prescription.  The differences arise from spatial variation of the background dipole field (not included in previous kinematic approaches) as well as from field evolution - the toroidal field reaches $\sim$60G at depth (consistent with $B_{\phi}\sim R_{m}B_{p}$) and the induced poloidal field reaches values three times larger than the imposed poloidal field, in contradiction of the simplifying assumptions made in previous models, see Figure 3.  

Overall, our 5G model wind speeds are decelerated 10-40\% compared to their non-magnetic counterparts (depending on latitude and radius), our 15G field shows similar slowing, except near the surface at high latitudes (Figure 2e shows the average of the absolute value of the zonal flow as a function of pressure for three different latitude ranges).  

\subsection{Ohmic Heating}
Contrary to previous results \citep{bat10,perna10b,bat11,rm13}, we find ohmic heating to be orders of magnitude too small to explain the inflated radius of HD209458b.  Figure 4 shows the ohmic dissipation rate calculated from the MHD model at 0.5 Bar and 100 Bar (a,c) and using the prescription outlined in Equation (6), with velocities from these simulations and a constant 5G field (b,d).  Similar to the Lorentz force, the prescription overestimates the maximum value of the ohmic dissipation at a give height by more than an order of magnitude.  We also see that Ohmic dissipation calculated self-consistently in these simulations is more localized than the prescription given by Equation (6).  Therefore, on integration over a given radius the ohmic dissipation will be substantially smaller than that given by the prescription.  This is true regardless of the windspeeds.  The integrated ohmic dissipation rate within a pressure scale height as a function of pressure for out hot model with two different magnetic field strengths is shown in (e).  Using a stellar insolation of 3.3$\times 10^{22}$W and the nominal required heating \citep{guillot02} of $\sim$ 0.1\% deep within the interior ($\sim$200 bar), one would require $\sim 10^{19}$W.  In Figure 4e we see that the peak ohmic dissipation occurs high in the atmosphere where it can not impact radius inflation.  Deeper in the atmosphere the measured ohmic dissipation is two orders of magnitude too small to explain inflation.  \cite{bat10} suggest a lower value of $\sim 10^{18}$W at 90bar could explain the radius of HD209458b, our models still show dissipation rates an order of magnitude too small.  The above results are consistent with the results of \cite{huangcumming12} who also found Ohmic dissipation was 1-2 orders of magnitude too small to explain the inflated radius of HD209458b.  Their model, which includes the back reaction of the field on the flow in determining ohmic dissipation rates, is closer to these MHD simulations than those of \cite{bat11} or \cite{wu13}.
%When comparing to \cite{rm13} we note that they use a model for HD209458b which is hotter than ours which could lead to different amplitudes of ohmic dissipation.  However, what we have showed in Figure 2 and 3 is that the {\it prescription} is wrong.  Furthermore, hotter models will have much higher values of magnetic Reynolds numbers, further invalidating the approach.  

\section{Discussion}
We have presented MHD simulations of the atmosphere of the hot Jupiter HD209458b.  We have found that magnetic fields slow winds by 10-40\% and contribute to ohmic heating of $\sim 10^{17}$W at depth.  Therefore, we find that ohmic dissipation, at least in the atmosphere modeled here, is unable to explain the inflated radius of this planet.  We also do not find any evidence that pole-ward meridional flows are arrested due to field geometry, as suggested in \cite{bat13}.  This is because our hot model is not hot enough to be in a magnetically dominated regime as assumed by \cite{bat13}.

Previous prescriptions treating magnetism in hot Jupiter atmospheres use a parameterization for the electric current based on Ohm's Law.  However, that current must also satisfy Ampere's law and Faraday's Law requires the time rate of change of the magnetic field be determined by the curl of the electric field.  That is, the solution to the full MHD equations is required to obtain a physically self-consistent representation of the appropriate fields and forces;  a parameterization of Ohm's Law alone is not sufficient.  Stated simply, a kinematic approach to MHD in these hot atmospheres does not adequately represent magnetic effects.  In the original work by \cite{perna10a} it was argued that the magnetic Reynolds number was generally less than one, hence validating their kinematic approach.  However, in hotter atmospheres (precisely those being investigated for ohmic inflation), the magnetic Reynolds number can be substantially larger than one.  In the hot model presented here (which is admittedly rather cool), the magnetic Reynolds number reaches values $\sim 10$, for hotter atmospheres of interest for ohmic heating, the Reynolds number could easily exceed $10^{3}$.  

There are several shortcomings of the models presented which we address here.  First, our reference state model is not realistic.  The cool model of \cite{iro05}, is known to be too cool and it shows virtually no magnetic effects.  Our adhoc ``hot'' model is also not physical.  However, such variations would not bring our models into better alignment with previous estimates.  Furthermore, our radiative transfer is crude, however, for these first MHD simulations, this heating is adequate and more realistic radiative transfer is not likely to alter our main results.  Another shortcoming of these models is that they are anelastic, so they do not accurately represent supersonic flows.  The comparisons shown in Figures 2 and 4, use the flow speeds in these MHD models, and we show that it is the details of the magnetic field that are inadequate, this would be true regardless of the flow speed.  However, larger flow speeds at depth could increase the value of the ohmic dissipation there, and therefore deep jets may be required to inflate planets.  In order to close the $\sim$2 order of magnitude gap in ohmic dissipation one would need MHD wind speeds an order of magnitude larger at depths greater than 10Bar.  While such wind speeds may be possible in purely hydrodynamic models, it is not clear whether they are likely in the presence of magnetic fields.  Additionally, our artificial boundary conditions could affect the profile of the radial current and therefore, the heating rate at depth.

Finally, probably the most problematic of the approximations we have made is that the magnetic diffusivity is only a function of the reference state temperature, instead of being a function of all space.  This is important because the stellar insolation changes the surface temperature substantially leading to dayside-nightside temperature variations which would cause large horizontal variations in the magnetic diffusivity.  This is likely to change the surface wind structure.  Orders of magnitude variation in magnetic Reynolds number could lead to magnetic instabilities which could lead to turbulence and mixing, which may affect horizontal temperature variations.  At the moment, it is unclear how these effects would alter wind structure.  We are currently implementing a spatially dependent magnetic diffusivity into our MHD model and results will be forthcoming.   

Despite its effect on wind structure at the surface, we do not expect a horizontally varying magnetic diffusivity to significantly alter our results on ohmic dissipation.  The stellar insolation does not effect the deeper layers much and therefore, the magnetic diffusivity will vary little from their reference state values.  Since it is those deeper layers that require heating in order to arrest contraction and lead to inflated radii, and those radii are affected the least by a horizontally dependent magnetic diffusivity we expect our result for ohmic dissipation to be robust (although see the caveats above with regard to wind speeds at depth).  Hotter internal temperatures may increase ohmic heating but those models will have higher $R_{m}$ and require substantially more computational resources.  Such models are currently being investigated.

\bibliographystyle{apj}

\begin{thebibliography}{44}
\expandafter\ifx\csname natexlab\endcsname\relax\def\natexlab#1{#1}\fi

\bibitem[{Baraffe {et~al.}(2003)Baraffe, Chabrier, Barman, Allard, \&
  Hauschildt}]{ba03}
Baraffe, I., Chabrier, G., Barman, T.~S., Allard, F., \& Hauschildt, P.~H.
  2003, Astronomy and Astrophysics, 402, 701

\bibitem[{Batygin {et~al.}(2013)Batygin, Stanley, \& Stevenson}]{bat13}
Batygin, K., Stanley, S., \& Stevenson, D.~J. 2013, The Astrophysical Journal,
  776, 53

\bibitem[{Batygin {et~al.}(2011)Batygin, Stevenson, \& Bodenheimer}]{bat11}
Batygin, K., Stevenson, D., \& Bodenheimer, P. 2011, The Astrophysical Journal,
  738, 1

\bibitem[{Batygin \& Stevenson(2010)}]{bat10}
Batygin, K., \& Stevenson, D.~J. 2010, The Astrophysical Journal Letters, 714,
  L238

\bibitem[{Bodenheimer {et~al.}(2003)Bodenheimer, Laughlin, \& Lin}]{bod03}
Bodenheimer, P., Laughlin, G., \& Lin, D. N.~C. 2003, The Astrophysical
  Journal, 592, 555

\bibitem[{Bodenheimer {et~al.}(2001)Bodenheimer, Lin, \& Mardling}]{bod01}
Bodenheimer, P., Lin, D. N.~C., \& Mardling, R.~A. 2001, The Astrophysical
  Journal, 548, 466

\bibitem[{Borucki {et~al.}(2009)Borucki, Koch, Jenkins, Sasselov, Gilliland,
  Batalha, Latham, Caldwell, Basri, Brown, Christensen-Dalsgaard, Cochran,
  DeVore, Dunham, Dupree, Gautier, Geary, Gould, Howell, Kjeldsen, Lissauer,
  Marcy, Meibom, Morrison, \& Tarter}]{borucki09}
Borucki, W.~J., {et~al.} 2009, Science, 325, 709

\bibitem[{Christensen {et~al.}(2001)Christensen, Aubert, Cardin, Dormy,
  Gibbons, Glatzmaier, Grote, Honkura, Jones, Kono, Matsushima, Sakuraba,
  Takahashi, Tilgner, Wicht, \& Zhang}]{christ01}
Christensen, U.~R., {et~al.} 2001, Physics of the Earth and Planetary
  Interiors, 128, 25

\bibitem[{Cooper \& Showman(2005)}]{coop05}
Cooper, C.~S., \& Showman, A.~P. 2005, Audio, Transactions of the IRE
  Professional Group on,

\bibitem[{Cowan \& Agol(2011)}]{cowan11}
Cowan, N.~B., \& Agol, E. 2011, The Astrophysical Journal, 729, 54

\bibitem[{Cowan {et~al.}(2007)Cowan, Agol, \& Charbonneau}]{cowan07}
Cowan, N.~B., Agol, E., \& Charbonneau, D. 2007, Monthly Notices of the Royal
  Astronomical Society, 379, 641

\bibitem[{Cowan {et~al.}(2012)Cowan, Machalek, Croll, Shekhtman, Burrows,
  Deming, Greene, \& Hora}]{cowan12}
Cowan, N.~B., Machalek, P., Croll, B., Shekhtman, L.~M., Burrows, A., Deming,
  D., Greene, T., \& Hora, J.~L. 2012, The Astrophysical Journal, 747, 82

\bibitem[{Crossfield {et~al.}(2010)Crossfield, Hansen, Harrington, Cho, Deming,
  Menou, \& Seager}]{crossfield10}
Crossfield, I. J.~M., Hansen, B. M.~S., Harrington, J., Cho, J. Y.-K., Deming,
  D., Menou, K., \& Seager, S. 2010, The Astrophysical Journal, 723, 1436

\bibitem[{Dobbs-Dixon \& Lin(2008)}]{iandd08}
Dobbs-Dixon, I., \& Lin, D. N.~C. 2008, The Astrophysical Journal, 673, 513

\bibitem[{Glatzmaier(1984)}]{glatz84}
Glatzmaier, G.~A. 1984, Journal of Computational Physics (ISSN 0021-9991), 55,
  461

\bibitem[{Glatzmaier(1985)}]{glatz85}
---. 1985, Astrophysical Journal, 291, 300

\bibitem[{Gough(1969)}]{go69}
Gough, D.~O. 1969, Journal of Atmospheric Sciences, 26, 448

\bibitem[{Guillot(2010)}]{guillot10}
Guillot, T. 2010, Astronomy and Astrophysics, 520, 27

\bibitem[{Guillot \& Showman(2002)}]{guillot02}
Guillot, T., \& Showman, A.~P. 2002, Astronomy and Astrophysics, 385, 156

\bibitem[{Heng {et~al.}(2011)Heng, Menou, \& Phillipps}]{heng11}
Heng, K., Menou, K., \& Phillipps, P.~J. 2011, Monthly Notices of the Royal
  Astronomical Society, 413, 2380

\bibitem[{Huang \& Cumming(2012)}]{huangcumming12}
Huang, X., \& Cumming, A. 2012, arXiv.org

\bibitem[{Iro {et~al.}(2005)Iro, B{\'e}zard, \& Guillot}]{iro05}
Iro, N., B{\'e}zard, B., \& Guillot, T. 2005, Astronomy and Astrophysics, 436,
  719

\bibitem[{Jones {et~al.}(2011)Jones, Boronski, Brun, Glatzmaier, Gastine,
  Miesch, \& Wicht}]{jones11}
Jones, C.~A., Boronski, P., Brun, A.~S., Glatzmaier, G.~A., Gastine, T.,
  Miesch, M.~S., \& Wicht, J. 2011, Icarus, 216, 120

\bibitem[{Knutson {et~al.}(2009{\natexlab{a}})Knutson, Charbonneau, Cowan,
  Fortney, Showman, Agol, \& Henry}]{knutson09b}
Knutson, H.~A., Charbonneau, D., Cowan, N.~B., Fortney, J.~J., Showman, A.~P.,
  Agol, E., \& Henry, G.~W. 2009{\natexlab{a}}, The Astrophysical Journal, 703,
  769

\bibitem[{Knutson {et~al.}(2007)Knutson, Charbonneau, Allen, Fortney, Agol,
  Cowan, Showman, Cooper, \& Megeath}]{knutson07}
Knutson, H.~A., {et~al.} 2007, Nature, 447, 183

\bibitem[{Knutson {et~al.}(2009{\natexlab{b}})Knutson, Charbonneau, Cowan,
  Fortney, Showman, Agol, Henry, Everett, \& Allen}]{knutson09a}
---. 2009{\natexlab{b}}, The Astrophysical Journal, 690, 822

\bibitem[{Knutson {et~al.}(2012)Knutson, Lewis, Fortney, Burrows, Showman,
  Cowan, Agol, Aigrain, Charbonneau, Deming, D{\'e}sert, Henry, Langton, \&
  Laughlin}]{knutson12}
---. 2012, The Astrophysical Journal, 754, 22

\bibitem[{Laughlin {et~al.}(2011)Laughlin, Crismani, \& Adams}]{laughlin11}
Laughlin, G., Crismani, M., \& Adams, F.~C. 2011, The Astrophysical Journal
  Letters, 729, L7

\bibitem[{Laughlin {et~al.}(2005)Laughlin, Wolf, Vanmunster, Bodenheimer,
  Fischer, Marcy, Butler, \& Vogt}]{laughlin05}
Laughlin, G., Wolf, A., Vanmunster, T., Bodenheimer, P., Fischer, D., Marcy,
  G., Butler, P., \& Vogt, S. 2005, The Astrophysical Journal, 621, 1072

\bibitem[{Lewis {et~al.}(2010)Lewis, Showman, Fortney, Marley, Freedman, \&
  Lodders}]{lewis10}
Lewis, N.~K., Showman, A.~P., Fortney, J.~J., Marley, M.~S., Freedman, R.~S.,
  \& Lodders, K. 2010, The Astrophysical Journal, 720, 344

\bibitem[{Maxted {et~al.}(2013)Maxted, Anderson, Doyle, Gillon, Harrington,
  Iro, Jehin, Lafreni{\`e}re, Smalley, \& Southworth}]{maxted13}
Maxted, P. F.~L., {et~al.} 2013, Monthly Notices of the Royal Astronomical
  Society, 428, 2645

\bibitem[{Menou(2012)}]{menou12}
Menou, K. 2012, The Astrophysical Journal, 745, 138

\bibitem[{Perez-Becker \& Showman(2013)}]{pbs13}
Perez-Becker, D., \& Showman, A.~P. 2013, The Astrophysical Journal, 776, 134

\bibitem[{Perna {et~al.}(2010{\natexlab{a}})Perna, Menou, \&
  Rauscher}]{perna10a}
Perna, R., Menou, K., \& Rauscher, E. 2010{\natexlab{a}}, The Astrophysical
  Journal, 719, 1421

\bibitem[{Perna {et~al.}(2010{\natexlab{b}})Perna, Menou, \&
  Rauscher}]{perna10b}
---. 2010{\natexlab{b}}, The Astrophysical Journal, 724, 313

\bibitem[{Rauscher \& Menou(2010)}]{rm10}
Rauscher, E., \& Menou, K. 2010, The Astrophysical Journal, 714, 1334

\bibitem[{Rauscher \& Menou(2013)}]{rm13}
---. 2013, The Astrophysical Journal, 764, 103

\bibitem[{Rogers(2011)}]{rogers11}
Rogers, T.~M. 2011, The Astrophysical Journal, 735, 100

\bibitem[{Rogers \& Glatzmaier(2005)}]{rg05}
Rogers, T.~M., \& Glatzmaier, G.~A. 2005, The Astrophysical Journal, 620, 432

\bibitem[{Showman {et~al.}(2009)Showman, Fortney, Lian, Marley, Freedman,
  Knutson, \& Charbonneau}]{show09}
Showman, A.~P., Fortney, J.~J., Lian, Y., Marley, M.~S., Freedman, R.~S.,
  Knutson, H.~A., \& Charbonneau, D. 2009, The Astrophysical Journal, 699, 564

\bibitem[{Showman \& Guillot(2002)}]{show02}
Showman, A.~P., \& Guillot, T. 2002, Astronomy and Astrophysics, 385, 166

\bibitem[{Showman \& Polvani(2011)}]{showpolv11}
Showman, A.~P., \& Polvani, L.~M. 2011, The Astrophysical Journal, 738, 71

\bibitem[{Wu \& Lithwick(2013)}]{wu13}
Wu, Y., \& Lithwick, Y. 2013, The Astrophysical Journal, 763, 13

\bibitem[{Youdin \& Mitchell(2010)}]{youdin10}
Youdin, A.~N., \& Mitchell, J.~L. 2010, The Astrophysical Journal, 721, 1113

\end{thebibliography}

\acknowledgments We are grateful to G. Glatzmaier, D. Lin, A.Cumming
and K. Batygin for helpful discussion and suggestions on improving the
manuscript.  Support for this research was provided by a NASA grant
NNG06GD44G.  T. Rogers is supported by an NSF ATM Faculty Position in
Solar physics under award number 0457631. A.P. Showman is supported by
NASA Origins grant NNX12A179G.

\end{document}